\documentclass[aps,prb,twocolumn,superscriptaddress,showpacs]{revtex4-1}
\usepackage{graphicx,bm,mathtools}

\begin{document}

\title{Superconductivity and multiple pressure-induced phases in BaPt$_2$As$_2$}

\author{C. Y. Guo}
\affiliation{Center for Correlated Matter and Department of Physics, Zhejiang University, Hangzhou, 310058, China}
\author{W. B. Jiang}
\affiliation{Center for Correlated Matter and Department of Physics, Zhejiang University, Hangzhou, 310058, China}
\author{M. Smidman}
\affiliation{Center for Correlated Matter and Department of Physics, Zhejiang University, Hangzhou, 310058, China}
\author{F. Han}
\affiliation{Materials Science Division, Argonne National Laboratory, Argonne, Illinois 60439, United States}
\affiliation{HPSynC, Geophysical Laboratory, Carnegie Institution of Washington, Argonne, Illinois 60439, United States}
\affiliation{Center for High Pressure Science and Technology Advanced Research, Beijing 100094, P. R. China}
\author{C. D. Malliakas}
\affiliation{Materials Science Division, Argonne National Laboratory, Argonne, Illinois 60439, United States}
\affiliation{Department of Chemistry, Northwestern University, Evanston, Illinois 60208, United States}
\author{B. Shen}
\affiliation{Center for Correlated Matter and Department of Physics, Zhejiang University, Hangzhou, 310058, China}
\author{Y. F. Wang}
\affiliation{Center for Correlated Matter and Department of Physics, Zhejiang University, Hangzhou, 310058, China}
\author{Y. Chen}
\affiliation{Center for Correlated Matter and Department of Physics, Zhejiang University, Hangzhou, 310058, China}
\author{X. Lu}
\affiliation{Center for Correlated Matter and Department of Physics, Zhejiang University, Hangzhou, 310058, China}
\affiliation{Collaborative Innovation Center of Advanced Microstructures, Nanjing 210093, China}
\author{M. G. Kanatzidis}
\affiliation{Materials Science Division, Argonne National Laboratory, Argonne, Illinois 60439, United States}
\affiliation{Department of Chemistry, Northwestern University, Evanston, Illinois 60208, United States}
\author{H. Q. Yuan}
\email{hqyuan@zju.edu.cn}
\affiliation{Center for Correlated Matter and Department of Physics, Zhejiang University, Hangzhou, 310058, China}
\affiliation{Collaborative Innovation Center of Advanced Microstructures, Nanjing 210093, China}

\date{October 24, 2016}

\begin{abstract}
The newly discovered  BaPt$_2$As$_2$ shows a structural distortion at around 275~K, followed by the emergence of superconductivity at lower temperatures. Here we  identify the presence of charge density wave (CDW) order at room temperature and ambient pressure using single crystal x-ray diffraction, with both a superlattice and an incommensurate modulation, where there is a change of the superlattice structure below $\simeq$ 275~K. Upon applying pressure,  BaPt$_2$As$_2$ shows a rich temperature-pressure phase diagram with multiple pressure-induced transitions at high temperatures, the emergence or disappearance of which are correlated with sudden changes in the superconducting transition temperature $T_c$. These findings demonstrate that BaPt$_2$As$_2$ is a promising new system for studying competing interactions and the relationship between high-temperature electronic instabilities and superconductivity.
\end{abstract}


\maketitle

\section{Introduction}
The presence of competing interactions and multiple  electronic instabilities often leads to emergent phenomena and new phases. In particular, the complex relationship between superconductivity and magnetic or charge order has attracted considerable interest. In  many of the high temperature iron pnictide superconductors, the  superconductivity occurs in the vicinity of spin-density-wave (SDW) order \cite{IronRev,IronRev2}. Meanwhile upon applying pressure to the heavy fermion superconductor CeCu$_2$Si$_2$ \cite{CCSA},  evidence was found for superconductivity being in close proximity to multiple instabilities, with one superconducting dome in the vicinity of magnetic order and another near a possible valence instability \cite{CCS1998,CCSB,CCSC}.  The interplay between charge-density-wave (CDW) order and superconductivity has also been of particular interest recently \cite{CDWsum1,CDW+SC2}, mainly due to the coexistence of these competing phases in some systems, with a similar phase diagram to systems with SDW and superconductivity \cite{CuTiSe}, as well as the observation of CDW in the high temperature cuprate superconductors. In the cuprates, the role of CDW order in the formation of superconductivity remains a central and open issue \cite{cupratesa,cupratesb,cupratesc,cuprates1,cuprates2,cuprates3}, but the microscopic relationship is still unresolved.\\
\begin{figure*}[t]
\begin{center}
  \includegraphics[trim = 0 80 0 130, clip, width=1.6\columnwidth]{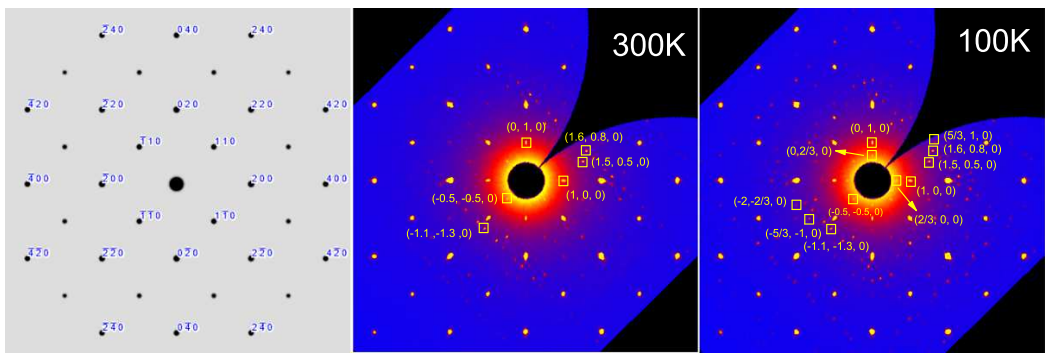}
\end{center}
	\caption{Reciprocal space reconstruction of single crystal XRD.  (a) Simulated diffraction peaks in the (hk0) plane of reciprocal space for the CaBe$_2$Ge$_2$-type structure. (b) Reciprocal space reconstruction of the single crystal XRD data collected at 300 K and (c) 100 K. At 300 K other than the simulated peaks of the CaBe$_2$Ge$_2$-type structure there are satellite peaks which can be indexed with a $2\times$2$\times2$ superlattice and an incommensurate modulation with a propagation vector  $\mathbf{q_1}$ = (0.1, 0.3, 0). At 100 K additional satellite peaks corresponding to a $6\times$6$\times2$ superlattice are observed.
}
   \label{XRDFig}
\end{figure*}

Structural distortions have also been found to occur in various iron based superconductors  \cite{Ba122SC}. For instance, upon lowering the temperature of electron doped BaFe$_2$As$_2$, the symmetry of the crystal lattice is reduced from the  four-fold  symmetry of the ThCr$_2$Si$_2$ structure, to a two fold symmetric orthorhombic structure, while SDW ordering onsets at lower temperatures \cite{Ba122edop,Ba122edop2}. There is an increasing body of evidence that this distortion does not arise due to phonons \cite{Ba122nophon},  but is from an electronic instablity, the nematic order, which may be driven by either  spin or orbital/charge fluctuations \cite{IronSCNemRev}. The suppression of both the nematicity and SDW ordering in the region of the superconducting dome, again indicates the importance of understanding the relationship between the various ordered phases and underlying interactions. On the other hand, there also exist some non-iron based 122 pnictide superconductors $RT_2Pn_2$ ($R=$Ba or Sr, $T=$transition metal and $Pn=$pnictogen) which undergo structural distortions upon cooling \cite{BaNiAs,SrPtSb,SPA SC}. A comparison with the iron based materials may provide crucial insights into the roles of various interactions and the resulting collective phenomena.

\begin{figure}[h]
\begin{center}
  \includegraphics[width=0.85\columnwidth]{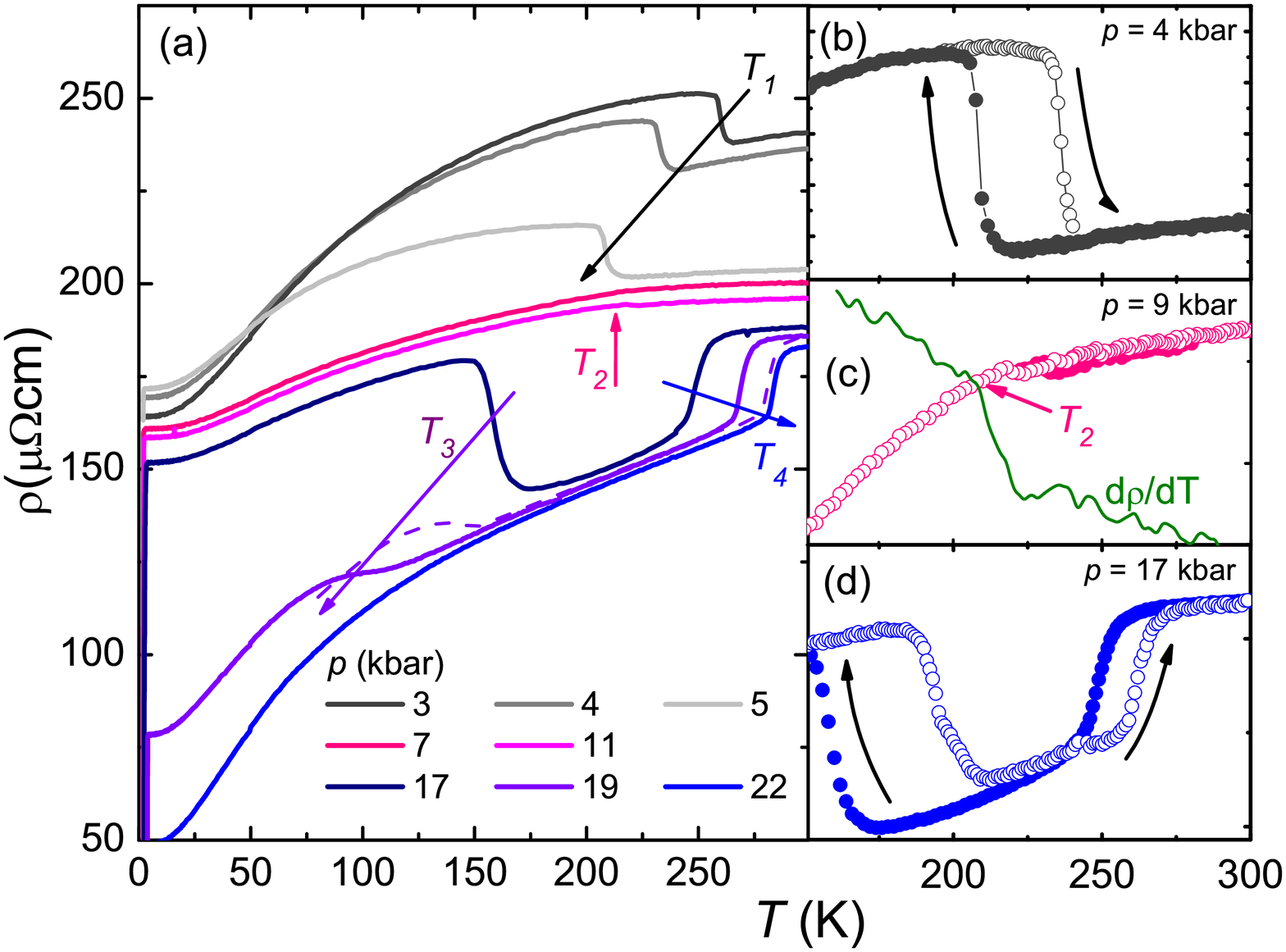}
\end{center}
	\caption{(a) Temperature dependence of the resistivity of BaPt$_2$As$_2$ at various pressures. The different high temperature transitions are denoted by $T_1$, $T_2$, $T_3$ and $T_4$. The resistivity upon warming and cooling is shown in the vicinity of (b) $T_1$ at 4~kbar,  (c) $T_2$ at 9~kbar, and (d) $T_3$ and $T_4$ at 17~kbar. In (c) the derivative $\rm{d}\rho/\rm{d}T$ is also shown by the solid line, which shows a clear change at $T_2$. }
   \label{Fig2}
\end{figure}

Unlike the 122 iron pnictide superconductors which crystallize in the ThCr$_2$Si$_2$ structure ($I4/mmm$), SrPt$_2$As$_2$ forms in the CaBe$_2$Ge$_2$-type structure ($P4/nmm$) at high temperatures. Both are layered tetragonal structures but they have different stacking of the layers of transition metal and As atoms along the $c$ axis.  A further difference from the iron pnictides is that in the Pt based materials, spin fluctuations are not generally expected to play a significant role. Below $T_{CDW}$~=~470~K, the crystal structure of  SrPt$_2$As$_2$ undergoes an orthorhombic distortion and CDW ordering \cite{Struc old,SPA R,SPA TEM,SPA NMR}, with a superconducting transition at a lower temperature of 5.2~K\cite{SPA SC}. Similarly in the newly discovered BaPt$_2$As$_2$ with a resistive transition at $T_c$~=~1.67~K \cite{BPAJWB}, a first order phase transition appears at about 275~K, at which a structural phase transition from the tetragonal  CaBe$_2$Ge$_2$-type structure to an orthorhombic structure was found using powder x-ray diffraction (XRD) \cite{BPAJWB}. In this article, we demonstrate using single crystal XRD  measurements that a CDW state already exists at room temperature in BaPt$_2$As$_2$ and  the structural transition at $T_1 \simeq 275$ K corresponds to a change in the CDW superlattice. To study the relationship between superconductivity and the CDW state, we performed resistivity measurements under pressure, which reveal a rich temperature-pressure phase diagram with multiple pressure-induced transitions, the emergence of which are correlated with sudden changes in $T_c$.

\section{Experimental Methods}
Single crystals of BaPt$_2$As$_2$  were synthesised using a self-flux method described in Ref~\onlinecite{BPAJWB}. Resistivity measurements under pressure were carried out using a piston-cylinder-type pressure cell up to 27~kbar, and Daphne 7373 was used as the pressure transmitting medium, to ensure hydrostaticity. The applied pressure was determined by the shift in $T_c$ of a high quality Pb single crystal.  Measurements were performed using a Physical Property Measurement System (Quantum Design PPMS-14T) in the temperature range of 2 K to 300 K, while measurements from 0.3~K to 5~K were performed using a $^3$He refrigerator.  Single-crystal XRD from room temperature to 100 K was carried out on a STOE IPDS II diffractometer at the Argonne National Laboratory, using an Mo source ($\lambda~=~0.71073$~\AA). Reciprocal space reconstruction of the single crystal XRD data was performed with the software X-Area.

\section{Results}
\subsection{Single crystal x-ray diffraction}

A simulation of the diffraction peaks in the ($hk0$) plane for the tetragonal CaBe$_2$Ge$_2$-type structure  ($P4/nmm$) is shown in  Fig.\ref{XRDFig}(a), while  the reciprocal space reconstruction of single crystal XRD data at 300~K and 100~K are shown in  Figs.\ref{XRDFig} (b) and (c) respectively. Reflections are labeled with the  Miller indices of the CaBe$_2$Ge$_2$-type structure ($P4/nmm$).

 In the previous powder XRD measurements \cite{BPAJWB}, we reported a structural phase transition at 275~K from a tetragonal CaBe$_2$Ge$_2$-type structure to an orthorhombic one. Based on the current single crystal XRD results, it should be noted that neither the reflections at 300~K nor 100~K could be satisfactorily indexed with a perfect CaBe$_2$Ge$_2$-type crystal structure. From comparing the experimental  data at both 300~K and 100~K with the simulated reflections, we observe a number of additional peaks arising from complex CDW modulations. If we index the single crystal XRD data by introducing an orthorhombic distortion, the lattice parameters at 300~K are $a~=~4.5608(10)~$\AA , $b~=~4.5643(12)~$\AA~and $c~=~10.0420(20)~$\AA, while at 100~K, $a~=~4.5489(9)~$\AA, $b~=~4.5700(11)~$\AA~ and $c~=~10.0192(22)~$\AA. These results indicate that even at room temperature, the CDW modulations induce a small difference in $a$ and $b$, which becomes larger below 280~K. Subsequent analysis of the data at 300~K reveals peaks corresponding to 2$\times$2$\times$2 superlattice structure, where the unit cell is doubled in all directions, along with an incommensurate modulation $\mathbf{q_1}~=~(0.1,0.3,0)$. Reflections which are forbidden for the CaBe$_2$Ge$_2$-type structure such as $(1,0,0)$ and $(0,1,0)$ and the half fractional peaks such as $(-1/2,-1/2,0)$ arise from the 2$\times$2$\times$2 superlattice structure while the peaks at the position of $(0.1,0.3,0)$ around the superlattice peaks arise from the incommensurate modulation. A comparison of the data from 300 K to 285 K shows no change in the structure of BaPt$_2$As$_2$. From 280 K to 100 K, BaPt$_2$As$_2$ enters a new phase in which the symmetry further lowers to a 6$\times$6$\times$2 superlattice structure while the incommensurate modulation $\mathbf{q_1}$  remains the same, as indicated by the  appearance of new fractional peaks such as $(2/3,0,0)$ and $(0,2/3,0)$. Therefore these results demonstrate that the CDW state is already present at room temperature, but there is a change in CDW structure below 280~K, along with an increased orthorhombic distortion.\\

\begin{figure}[t]
\begin{center}
 \includegraphics[width=0.85\columnwidth]{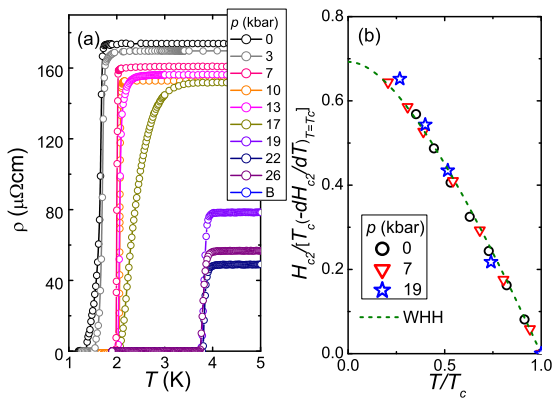}
\end{center}
\caption{(Color online)(a) Resistivity of BaPt$_2$As$_2$  at various pressures, near the superconducting transition. (b) The upper critical field ($H_{c2}$) of BaPt$_2$As$_2$ as a function of $T/T_c$ at various pressures, normalized by the product of $T_c$ and the slope of $H_{c2}$ near $T_c$.}
\label{Fig3}
\end{figure}
\subsection{Resistivity under pressure}
The temperature-pressure phase diagram of BaPt$_2$As$_2$ was determined by measuring the resistivity under pressure. As shown in Fig.~\ref{Fig2}(a), at $p~=~3$~kbar, there is a sudden increase of the resistivity upon cooling at about $T_1\simeq 250$~K, which is similar to the previously reported results at ambient pressure\cite{BPAJWB} and corresponds to a change of the CDW structure. With increasing pressure, the temperature of this transition decreases, before abruptly disappearing above $p_{1}\simeq7$~kbar. After $T_1$ disappears, a new transition labelled as $T_2$ is observed, where there is only a small anomaly in the resistivity and little change with increasing pressure. 
 For pressures greater than  $p_{2}\simeq17$~kbar, $T_2$ disappears and two new transitions emerge at $T_3$ and $T_4$, with the former corresponding to an increase of the resistivity upon cooling while the latter shows a  sudden decrease.  These transitions have  a different pressure dependence with the lower transition $T_3$ being rapidly suppressed with increasing pressure, while $T_4$ increases with pressure. Note that apart from $T_2$, all the transitions display pronounced hysteresis, as shown in Figs.\ref{Fig2}(b)-(d), indicating a  first-order nature. It can also be seen that there is an increase in  resistivity at $T_1$ and $T_3$ while it drops at $T_4$. 

\begin{figure}[t]
\begin{center}
 \includegraphics[width=0.75\columnwidth]{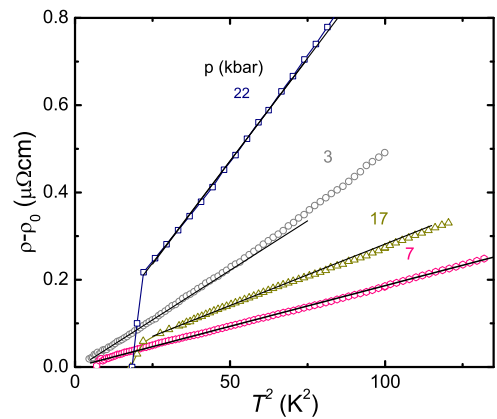}
\end{center}
\caption{(Color online) Resistivity as a function of $T^2$ at several pressures with the residual values of $\rho_0$ subtracted. The solid lines show fits to a $\sim T^2$ dependence in the normal state.}
\label{Fig4}
\end{figure}

Figure~\ref{Fig3}(a) displays the resistivity under pressure at low temperatures, in the vicinity of the superconducting transitions. At ambient pressure, $T_c$ is about 1.6~K, determined from where the resistivity reaches $50\%$ of the normal state value $\rho_0 $$\simeq$$ 174 ~\mu\Omega$cm, similar to that previously reported\cite{BPAJWB}. With increasing pressure, the superconducting transition sharpens and $T_c$ undergoes a moderate enhancement at around 7~kbar which is close to $p_1$, to around 2~K, along with a small decrease of $\rho_0$. Further increasing the pressure up to 13~kbar results in little change of $T_c$. However, upon applying pressures greater than 13~kbar, there is another jump in $T_c$ which is larger in size, reaching a near constant value above 17~kbar. The superconducting transition at 17~kbar is significantly broader than that at both lower and higher pressures, which may be due to a pressure inhomogeneity, leading to the coexistence of both superconducting phases. The newly emerged superconducting phase has the highest $T_c$ $\simeq$ 3.8 K, the lowest residual resistivity in the normal state, and $T_c$ changes little with pressure up to 26~kbar. The upper critical field $H_{c2}(T)$ at different pressures is shown in  Fig.~\ref{Fig3}(b). The values of $H_{c2}$ were measured in three regions of the temperature-pressure phase diagram (0, 7 and 19~kbar), and were normalized by the product of $T_c$ and the derivative of $H_{c2}$ near $T_c$. When the normalized $H_{c2}$ is plotted as a function of $T/T_c$, all the data falls on one curve, in good agreement with the orbital limited values of $H_{c2}$ calculated using the Werthamer-Helfand-Hohenberg (WHH) model shown by the dashed line \cite{WHH3}. This suggests that there is no change of the superconducting pairing state under pressure and that orbital limiting rather than Pauli limiting is the dominant pair breaking mechanism. Meanwhile as displayed in Fig.~\ref{Fig4}, the low temperature normal state resistivity could be well fitted with Fermi liquid behavior, $\rho(T)~=~\rho_0+AT^2$.\\

\begin{figure}[t]
\begin{center}
 \includegraphics[width=0.9\columnwidth]{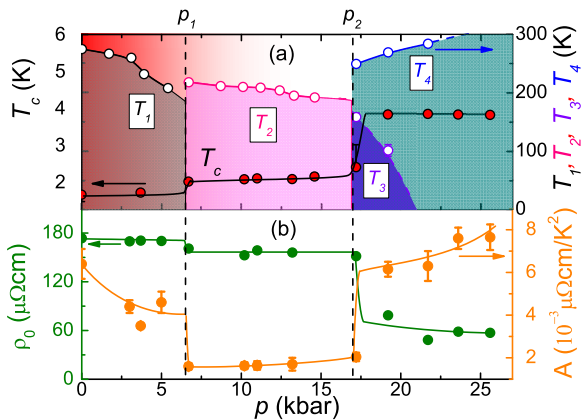}
\end{center}
\caption{(Color online) (a) Temperature-pressure phase diagram of  BaPt$_2$As$_2$ showing both the high temperature and superconducting transitions. (b) Pressure dependence of the residual resistivity $\rho_0$ and $A$ coefficient.}
\label{Fig5}
\end{figure}

Figure~\ref{Fig5}(a) displays the temperature-pressure phase diagram obtained from resistivity measurements. Two abrupt enhancements of $T_c$ occur under pressure, both of which clearly coincide with changes in the high temperature phase transitions. 
At $p_{1}$ $\simeq$ 7~kbar, the CDW transition at $T_1$ disappears and a new transition  $T_2$ is observed. This change is accompanied by an enhancement of $T_c$. Upon further increasing the applied pressure, there is little change of $T_2$ up to $p_2 \simeq 17$~kbar, where the transition at $T_2$ is no longer observed and two new phase transitions emerge. The transition at  $T_3$  decreases with increasing pressure before being suppressed at less than 22 kbar, while $T_4$ increases with pressure, reaching 290~K at 24~kbar. The appearance of these transitions is accompanied by  another pronounced enhancement of $T_c$ by almost a factor of two. The critical pressure for the emergence of $T_3$ and $T_4$ is slighty lower than that corresponding to the jump in $T_c$. This may be due to the variation in the pressure transmitted by the Daphne~7373 medium with temperature, which can cause a pressure increase of around 1-2~kbar between 300~K and low temperatures \cite{DaphRef}. Furthermore, the results from fitting the resistivity in the normal state are shown in Fig.~\ref{Fig5}(b). Each step like increase of $T_c$ coincides with a drop in $\rho_0$, with a small decrease at the disappearance of $T_1$ and a larger drop at the emergence of $T_3$ and $T_4$. Meanwhile the resistivity coefficient $A$ undergoes a sharp drop at $p_1$ followed by a sudden increase at $p_2$. Furthermore, neither $\rho_0$ or $A$ display any noticeable anomaly  upon the disappearance of the transition at $T_3$.

\section{Discussion and summary}

The above results demonstrate an intricate relationship between the superconductivity and the high temperature transitions in BaPt$_2$As$_2$. The nature of the transitions at $T_2$, $T_3$ and $T_4$ are yet to be uncovered.  In comparison, SrPt$_2$As$_2$ undergoes a structural phase transition below $T_{CDW}~=~470$~K, from a tetragonal to orthorhombic structure \cite{Struc old}, with two incommensurate CDW modulations emerging \cite{SPA R, SPA TEM}. However, in BaPt$_2$As$_2$ at ambient pressure  the orthorhombic distortion and CDW modulation already exist above  $T_1$ and this transition corresponds to a change of the periodicity of the superlattice. Furthermore, there is a sudden increase in the resistivity at $T_1$ as in the case of the structural transition in SrNi$_2$P$_2$  \cite{SrNiP}, whereas there is a drop in the resistivity at  $T_{CDW}$ in SrPt$_2$As$_2$. However, the behaviour of BaPt$_2$As$_2$ above 22~kbar more strongly resembles that of  SrPt$_2$As$_2$, since we observe a drop in the resistivity at $T_4$ and there is  a comparable value of  $T_c$ .

If then the transition under pressure at $T_4$ is similar to that of SrPt$_2$As$_2$ at ambient pressure, the differences in the CDW modulation and high temperature crystal structures may explain the different behavior of the resistivity seen at $T_1$ and $T_4$. This suggests the possibility that the CDW state below $T_1$ disappears at $p_1$, above which only a weak anomaly at $T_2$ is observed.  Furthermore, it is not clear whether the CDW state observed at ambient pressure and room temperature persists upon the application of pressure and it may be that the transition at $T_2$ lies beneath a CDW transition at higher temperatures.  At $p_2$ a different CDW state may emerge. However, while CDW states generally compete with superconductivity, in this instance the emergence of the transition at $T_4$ leads to a significant enhancement of $T_c$, rather than a decrease. The dramatic change in both $\rho_0$ and the $A$ coefficient at $p_2$ suggests this may correspond to a sudden change in the electronic structure and Fermi surface topology. Such changes may lead to an enhanced density of states, which in turn may cause the significant increase of $T_c$ and $A$. 

The transition at $T_3$, which appears under $T_4$ up to around 19~kbar, may then correspond to a further lowering of the crystal symmetry or a change in the modulation vector  \textbf{q}  of the CDW state. A change of \textbf{q} can lead to different behavior of the resistivity since this changes the  reconstructed Fermi surface, with the CDW gap opening across different regions. However for these scenarios, it would be unusual that there is  little change in $T_c$  upon the disappearance of $T_3$.  As previously discussed, at ambient pressure there is a  structural distortion at $T_1$ corresponding to a significantly increased orthorhombicity.  Since both $T_3$ and $T_4$ have a similar first order nature, whether these transitions correspond to changes in either the crystal symmetry or structural parameters needs to be checked with XRD measurements under pressure. All these results suggest the presence of both structural and electronic instabilities in the temperature-pressure phase diagram. In light of the recent evidence that the nematic order in the iron pnictide based superconductors is driven by electronic rather than structural instabilities \cite{IronSCNemRev}, it will be of interest to determine whether any of the transitions or structural changes in BaPt$_2$As$_2$ also have an electronic origin. The relationship between the high temperature transitions and the superconductivity also needs to be further investigated and compared to both the cuprate and iron based materials, but the significantly lower values of $T_c$ may be a result of  the lack of a spin fluctuation mediated pairing mechanism.

In conclusion,  using single crystal XRD measurements we have identified that a CDW state exists in BaPt$_2$As$_2$ in ambient conditions, and that the transition at $T_1~=~275$~K corresponds to  a change of the superlattice structure. The application of pressure results in a rich phase diagram. Pressure suppresses the CDW transition at $T_1$ until it disappears at around $p_1$~=~7~kbar, which leads to an enhancement of $T_c$. Upon further increasing the pressure,  new high temperature transitions are observed in the resistivity, which are correlated with an additional increase of $T_c$. Since these results indicate the presence of multiple instabilities at high temperatures,  BaPt$_2$As$_2$ may provide a platform for studying the competition between different interactions and their interplay with superconductivity. Further experiments such as NMR and XRD under pressure, as well as theoretical calculations are  needed to characterize the origin of these transitions and elucidate the mechanisms which lead to these exotic behaviors.

\begin{acknowledgments}
We thank C. Cao for interesting discussions. The work at Zhejiang University was supported by the National Natural Science Foundation of China (No.11474251 and No.11374257), National Key R\&D Program of the MOST of China (No. 2016YFA0300202) and the Science Challenge Program of China. The work at Argonne was supported by the Office of Science, Office of Basic Energy Sciences, Division of Materials Sciences, US Department of Energy, under contract DE-AC02-06CH11357 (BES-DMSE).
\end{acknowledgments}

\end{document}